# Are Deep Learning-Generated Social Media Profiles Indistinguishable from Real Profiles?

Sippo Rossi
Copenhagen Business School
sr.digi@cbs.dk

Youngjin Kwon
Temple University
youngjin.kwon@temple.edu

Odd Harald Auglend
Copenhagen Business School
oha.digi@cbs.dk

Raghava Rao Mukkamala
Copenhagen Business School
rrm.digi@cbs.dk

Matti Rossi
Aalto University
matti.rossi@aalto.fi

Jason Thatcher
Temple University
jason.thatcher@temple.edu

**Abstract**

*In recent years, deep learning methods have become increasingly capable of generating near photorealistic pictures and humanlike text up to the point that humans can no longer recognize what is real and what is AI-generated. Concerningly, there is evidence that some of these methods have already been adopted to produce fake social media profiles and content. We hypothesize that these advances have made detecting generated fake social media content in the feed extremely difficult, if not impossible, for the average user of social media.*

*This paper presents the results of an experiment where 375 participants attempted to label real and generated profiles and posts in a simulated social media feed. The results support our hypothesis and suggest that even fully-generated fake profiles with posts written by an advanced text generator are difficult for humans to identify.*

## 1. Introduction

Disinformation, conspiracy theories, links to phishing sites and even legitimate sales pitches are being sent with the help of fake social media profiles to unsuspecting users of social networking sites (Bond, 2022; Shafahi et al., 2016; Shao et al., 2018). These fake profiles might be automated bots belonging to a network of such or be operated by a human that manages multiple accounts. Previously maintaining realistic avatars and producing convincing content at scale was difficult as stolen images could be found with a reverse image search. Similarly, even the best generated text was unable to consistently pass for something written by a human.

As a result of recent rapid advances in deep learning (DL) methods that can be used to generate realistic images and due to the proliferation of advanced pre-trained natural language processing (NLP) models, creating synthetic profile pictures and producing human-like texts is easier than ever before (Brown et al., 2020; Köbis & Mossink, 2021; Nightingale & Farid, 2022). Consequently, producing large numbers of fake profiles with individual or even multiple synthetic components such as photorealistic profile pictures and generated but human-like posts is now technologically and economically viable. Thus, these advances may have made detecting sophisticated fake profiles near impossible for the average user of a social networking site. From the perpetrator's point of view, this has made information operations and trolling cheaper and much less labor-intensive. If these fake profiles are not distinguishable by humans and large quantities of them can be produced easily, we could see an explosion of bot accounts being used for marketing, phishing or political campaigns among other purposes.

It is not yet known whether fully synthetic profiles and social media posts are in fact able to pass the Turing test and go unnoticed by humans. Therefore, the primary goal of this paper is to study how well humans can detect deep learning-generated social media profiles and posts in a social networking site's feed, as well as to assess the ability of modern deep learning to produce humanlike content. Our first research question is:

**RQ1:** Can humans distinguish social media profiles with DL-generated profile pictures and DL-generated posts from real ones in the feed of a social networking site?

While it is known that in isolation these individual generated components are no longer distinguishable from real ones, there is a possibility that in some cases when combined within one profile they become suspicious. As an example, consider the situation when the text in a post contains vocabulary used by a certain demographic group, but the profile picture belongs to clearly different group. Therefore, our second research question is:

**RQ2:** Which components of a profile are more likely to make humans suspect that the profile is fake?

We hypothesize that we have crossed the boundary where generated social media profiles can no longer be consistently detected by humans. To test this hypothesis and to answer the research questions, we conducted an experiment where participants were shown both genuine and fully generated bot profiles in a simulated social media feed and asked to classify the accounts and assess different components of the profiles on whether they are suspicious or not. The feed contained both the basic profile information as well as one post made by each account.

This paper begins by briefly synthesizing the findings of recent literature related to image and text generation and fake content on social media. Next, the experiment and methodology used to produce the social media profiles are explained. We then describe the results of the experiment and discuss the main findings and implications. Lastly, we conclude by considering the limitations of this study and provide an overview of the planned future work that will address them.

## 2. Background and related research

In this section, we first briefly discuss recent trends in fake account detection. Then, we summarize the state of the art in image and text generation using deep learning methods and lastly review recent experiments involving fake social media content.

### 2.1. Bots and fake content on social media

During the recent years bots on social media as well as methods to detect them have been studied in both information systems (Ross et al., 2019; Salge et al., 2022; Stieglitz et al., 2017; Williamson & Scrofani, 2019) and computer science research (Cresci, 2020; Ferrara et al., 2016). There are also a number of studies on the prevalence and impact of bots under various topics such as elections (Brachten et al., 2017) or the COVID-19 pandemic (Marx et al., 2020; Rossi, 2022). Fake content such as misinformation and fake news (Kim & Dennis, 2019; Moravec et al., 2019), as well as the role of bots in distributing them have also been investigated (Lazer et al., 2018; Shao et al., 2018).

However, academic research on the human ability to detect deep learning-generated content is scarce, based on our search. We assume this is most likely due to the technology only recently having matured enough to produce believable fake content. Meanwhile, algorithmic detection of fake images and bot profiles on social media have been studied more, despite the relative recentness of the topic (Cresci, 2020; X. Wang et al., 2022; Yu et al., 2019). From other non-academic sources, there are documented examples of cases where fake profiles were caught using GAN images on Twitter (Strick, 2021) and LinkedIn (Bond, 2022) to pass as real humans. While some examples were more benign, groups of accounts with GAN images impersonating real humans have been said to be employed for promoting computational propaganda (Da San Martino et al., 2021; Strick, 2021).

### 2.2. Text and image generation using deep learning

Deep learning methods for text generation have become more accessible due to powerful pre-trained models such as GPT-2 and GPT-3 (Brown et al., 2020; Li et al., 2021). In the past language models required significant computational resources and large dataset sets to train them for each individual topic. In contrast, pre-trained models are trained with massive amounts of training data before being released to the public. Therefore, they do not have this limitation as they can be used immediately or after being fine-tuned with much more manageable datasets and computing power (Li et al., 2021).

These powerful pre-trained models for text generation have been researched and developed primarily by leading technology companies and private research organizations in recent years. Therefore, there is still a limited number of peer-reviewed academic research on, for example on, how well humans can distinguish human-written texts from the auto-generated text. Early works have shown that GPT-3 is, for example, capable of producing poems that are indistinguishable from genuine ones (Köbis & Mossink, 2021) and that humans have difficulties even after training to detect machine-generated text (Clark et al., 2021). It has also been suggested that GPT-2 has been used to produce texts for malicious accounts (Da San Martino et al., 2021), although it is difficult to prove and the effectiveness of it is still unknown.

For image generation, Generative Adversarial Networks (GANs), a type of deep learning architecture, have been demonstrated to be able to produce synthetic images that are algorithmically detectable, but for humans seemingly photorealistic (Karras et al., 2019; Yu et al., 2019). A recent experiment with facial images generated with StyleGAN (Karras et al., 2019), an advanced and alternative architecture for GANs, demonstrated how synthetic images were deemed on average more trustworthy than real faces and nearly undetectable (Nightingale & Farid, 2022). Figure 1 contains examples of GAN-generated images of human faces.

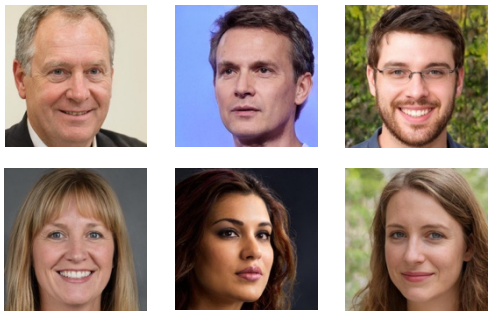

**Figure 1. GAN-generated images**

### 2.3. Experiments and fake content on social media

Experiments are a common method in studies related to deception and misinformation on social media. We identified two main approaches for experiments in social media, which are either conducting experiments directly within the social media platform (Cresci et al., 2017; Freitas et al., 2015; Shafahi et al., 2016) or alternatively by using a survey or other simulated environment such as a web-based game (Moravec et al., 2019; Roozenbeek & Linden, 2019).

While conducting the experiment with a simulated social media page can limit realism, they are generally less risky due to the environment being controlled and since debriefing is possible as well as acquiring informed consent from participants. Failure to take the appropriate measures has resulted in criticism of such research in the past (Flick, 2016). Therefore, for this study we preferred simulated environments to reduce potential ethical concerns.

Experiments with similar methods and goals to this study have been conducted in social bot research, where researchers have used crowdsourcing to determine whether humans can detect different types of bots, such as social bots and spambots (Cresci et al., 2017; G. Wang et al., 2013). The main difference in these studies was that the profiles of the bots were not generated using deep learning, and that the participants had access to view complete profiles, while in our study participants are shown a view similar to social media feeds. We argue that showing only what is visible in the feed is more realistic, as the average user might not go meticulously through each profile that they come across.

Other related experiments are the previously mentioned tests on the human ability to detect generated content in the context of poems (Köbis & Mossink, 2021) and profile pictures (Nightingale & Farid, 2022), which have shown that individual components similar to those in social media profiles can fool humans, but based on our knowledge no studies have yet at the time of writing been conducted on the human ability to detect fully deep learning-generated profiles.

### 3. Research design

In this section, we will explain the setup of the experiment and then describe the process used to generate the fake profiles and tweets, as well as how the real profiles and posts were collected.

The experiment imitated a situation where a Twitter user is scrolling through the feed and sees a post and the limited set of profile information about the account that posted it. The manipulations were built into pages hosted by Qualtrics using images of profiles along with the posts as shown in Figure 2. Participants were recruited using Amazon's Mechanical Turk (MTurk).

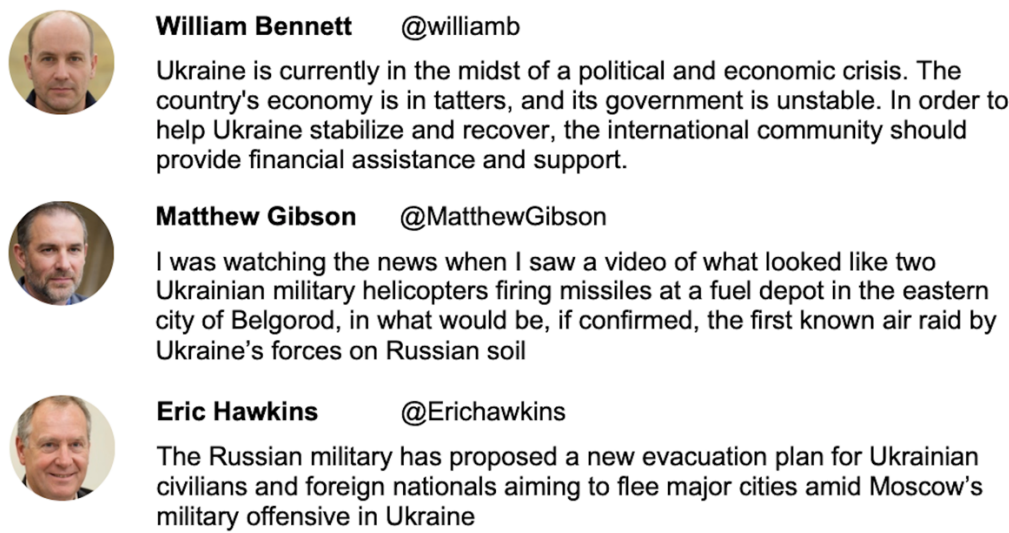

**Figure 2. Examples of the generated profiles shown during the experiment**

This simulated approach to having a mock Twitter feed was chosen for two reasons. First, this ensured that the auto-generated profiles and posts were not seen by any non-participating individuals. This could have been an issue as some of the generated content can be described as misinformation or otherwise controversial. Second, in this way we avoided violating the terms of use of Twitter and GPT-3, the deep learning-based language model used to generate texts of the posts.

The topic of the posts shown during the experiment was the war in Ukraine. It was chosen because of the timeliness of the topic and since a vast number of suitable real tweets were readily available from verified accounts. The style of the real tweets and accounts that were selected for the experiment are described in detail in the methodology section. While the posts discussed the war and sometimes contained questionable content, the experiment did not have any mentions of violence or other forms of graphic or disturbing content. Participants were also warned of the subject before being shown the posts and given an opportunity to back out without any consequences.

Due to the controversial nature of the topic and possible differences in views of participants, we designed the experiment to include real and generated profiles with tweets supporting both sides. During the analysis, we checked that the personal view of the participant did not correlate with how they rated perceived accounts as genuine or fake.

To reduce low-quality responses and to ensure the survey worked smoothly, we initially ran a limited trial experiment with approximately 100 subjects. We then improved the survey based on feedback received during trial. The experiment had a screening phase which was used to recruit US citizens that speak English fluently and that have experience with social networking sites, including Twitter. Lastly, the experiment was designed to be completed in a short time to reduce fatigue or learning effects. To incentivize participants to respond properly, they were instructed that the experiment includes an attention check and that identified cases of rushing through the survey or failing to respond adequately would result in no reward. The survey was terminated upon failure to answer correctly on the question containing the attention check. The attention check was placed near the beginning of the survey to not waste the time of inattentive participants. Ultimately all participants who passed the screening, passed the attention check and completed the survey were given the reward regardless of their performance.

During the experiment each participant was told that they are going to be shown profiles that belong to humans or bots, with the bots being the accounts that are deep learning-generated. The participants were shown four profiles, which were drawn from a total of 9 bots and 9 humans. For each profile shown, the participants were asked to label whether the account is a "bot" or a "human" and to rate their perceived likelihood of the account being a bot as well as to score different components in terms of whether they make the account seem suspicious, on a scale of 1 (not at all suspicious) to 10 (extremely suspicious).

We opted to have only white adult male profiles with common English names both in the real and generated profiles, to eliminate the influence of gender, race or perceived nationality on the results. Furthermore, fields such as the time of the post, likes and retweets were removed to control for their effects. While the number of comments, likes and retweets can have an impact on the credibility of a post, these can also be inflated with the use of bots or bought interactions from follower farms.

The following two sections describe in detail the process used to generate the fake profiles as well as how the real profiles were collected.

### 3.1. Generating fake profiles and tweets

The generated profiles and accompanying posts consisted of four elements, which were created using an automated script and with as little human intervention or tweaking as possible. This was done to imitate the mass production of fake accounts. The four generated elements were the profile picture, name, handle and post (tweet). Out of these, the profile picture and post were generated using deep learning-based methods and the name and handle with a basic script written in Python.

The profile pictures were scraped from the website "thispersondoesnotexist.com", which produces a unique image every time the page is visited using StyleGAN, an advanced generative adversarial network (GAN) model. This crude approach demonstrates how easy it is to get a large sample of fake images generated by deep learning. While the method is straightforward, the underlying GAN model itself is an innovative approach to image generation. GAN models consist of two separate neural networks, a generator that produces images and discriminator that attempts to classify real images from the synthetic given to it by the generator. The discriminator provides feedback to the generator, which adjusts its parameters until the produced images are no longer detected by the discriminator (Creswell et al., 2018).

The posts were generated using OpenAI's Generative Pre-trained Transformer 3 (GPT-3), which is an advanced deep learning-based NLP model that can be used among other things to produce high-quality text, based on prompts (Brown et al., 2020). GPT-3 has 175 billion parameters and has been trained with several massive datasets consisting of for example Wikipedia pages, text collected by crawling the internet and two large internet-based books corpora (Brown et al., 2020). As an autoregressive transformer model, given some input it can predict very accurately for example what words would complete a sentence or what is an appropriate response to a question.

While officially GPT-3 bans its use for generating any content including posts that are shown on social media, we applied and received an exemption to use it for this purpose in our experiment. The posts were created by prompting the model to summarize news articles related to the war in Ukraine with slight randomized variations in the parameters. Examples include asking the model to write the summary as a positive or negative opinion or by requesting that it explains the given text in language understandable to children. Due to the sensitive nature of the topic, we manually checked each text before including it in the experiment, to determine whether it contained any kind of violence or any other kind of objectionable content.

The two final and related fields shown in the experiment are the username and handle. The names consisted of a first and last name that were generated with a script that used a list of common US names. The handles were then derived from the names using a stochastic process that cut the first or last name to only the initial, and occasionally added random numbers to the end. After qualitatively inspecting the produced names and handles, they were deemed sufficiently similar to those of real Twitter users.

### 3.2. Collecting real profiles and tweets

The real profiles shown during the experiment mainly belonged to verified profiles of journalists, celebrities, pundits and politicians, whose identity and status as real humans could be easily verified and whose posts were public. The profiles included had to have a clear profile picture containing their face as well as full names so that they were comparable with the generated profiles and would not introduce any unnecessary noise to the experiment and influence the results.

The profiles were collected by retrieving verified profiles tweeting about Ukraine. These were then manually checked and included in the experiment if they met the criteria regarding the profile picture and name described above. Lastly, to reduce the chance of the real profiles being too well-known and thus recognizable by the participants, we removed profiles that belonged to very high-ranking politicians such as ministers or leaders of states as well as accounts which multiple authors could recognize.

## 4. Results

In this section, we first describe the demographics of the participants in the experiment and then examine the classification accuracies and assess the participants' ability to identifying the fake/real posts. Lastly, we discuss the perceived suspiciousness of the components according to the participants.

### 4.1. Participants

The results presented in this section are from the experiment held in May 2022 through MTurk. Out of 1292 subjects who participated in the screening, 478 were invited to complete the experiment. Ultimately, 375 participants both completed the experiment and passed the attention check at the beginning of the survey. The results discussed are only for the 375 subjects that successfully completed these steps. The average duration that it took to complete the experiment was 10 minutes and 44 seconds.

**Table 1. Demographics**

| Category | Sub-Category | N | % |
|---|---|---|---|
| Gender | Male | 210 | 56,0 % |
| Gender | Female | 164 | 43,7 % |
| Gender | Other | 1 | 0,3 % |
| Age | 19-29 | 41 | 10,9 % |
| Age | 30-39 | 132 | 35,2 % |
| Age | 40-49 | 104 | 27,7 % |
| Age | 50-59 | 51 | 13,6 % |
| Age | 60-69 | 39 | 10,4 % |
| Age | 70+ | 8 | 2,1 % |
| Race / Ethnicity | White | 313 | 83,5 % |
| Race / Ethnicity | Asian | 21 | 5,6 % |
| Race / Ethnicity | Hispanic | 19 | 5,1 % |
| Race / Ethnicity | Black | 11 | 2,9 % |
| Race / Ethnicity | Native American | 5 | 1,3 % |
| Race / Ethnicity | Other | 6 | 1,6 % |
| Highest Degree | Primary school | 1 | 0,3 % |
| Highest Degree | High school | 54 | 14,4 % |
| Highest Degree | Bachelor's | 258 | 68,8 % |
| Highest Degree | Master's | 57 | 15,2 % |
| Highest Degree | Doctoral | 5 | 1,3 % |

The participants predominantly identified as white (83.5%) and with a slight skew towards males, representing 56% of all subjects. The average age was 38 years and a most subjects (85.3%) have at least a bachelor's degree, meaning that the participants are more educated than the US population on average. The demographic information is summarized in Table 1. The lack of non-white participants can introduce bias to the results, making generalizing the findings to the general population risky. Therefore, these findings will be more relevant to assessing the capability of this particular demographic's ability to detect fake content.

To check that the experiment and survey's designs and instructions were sufficient, each participant had to rate both the clarity of the instructions and the perceived difficulty of the task after completing the survey on a 5-point Likert scale. When asked if "the tasks and instructions were clear", 96% responded either agree or strongly agree. When asked if "the given task was easy to do", the result was more spread with 83% stating that they agree or strongly agree, 11% neither agreeing nor disagreeing, and the remaining 6% disagreeing or strongly disagreeing. Based on these results the instructions were adequate. Despite the poor performance in terms of classification accuracy, only a small share of the subjects viewed the task difficult.

### 4.2. Classification accuracy and likelihood

Since each participant was shown 4 randomly drawn profiles, the number of times each profile was labeled during the experiment varied from 77 to 85. None of the eighteen profiles received unanimous labels and when inspecting the accuracy by profile (i.e., the percentage of time it was correctly labeled), for the generated accounts the accuracies ranged from 10% to up to 27.4%. The generated profiles had a mean of 18.2% and 95 percent confidence interval (CI) at [0.145, 0.219]. The genuine profiles had accuracies ranging from 58.5% to 91.4% with a mean of 79.7% and a 95 percent CI at [0.737, 0.856]. This suggests that the participants were unable to reliably detect the generated profiles. Interestingly, the most divisive accounts belonged to genuine humans. One of the real profiles was mislabeled by 41.5% of the participants that saw it. Meanwhile, the two best-performing fake profiles shown in Figure 3 were labeled as bots by only 10% and 11.1% respectively. The mean accuracy considering all profiles was 48.9%, which, given that there was an equal amount of genuine and generated profiles, is close to a random guess. The survey results are summarized in Table 2.

The participants were also asked to rate the likelihood of the account being a bot on a 5-Point Likert scale. This was done so that the level of certainty for the labeling could be assessed, with 1 being very unlikely and 5 very likely. The mean likelihood given to the generated profiles was 3.19 while for the genuine profiles it was 3.26. This indicates that the participants were on average uncertain of their labeling, regardless of whether accounts are bots or humans.

**Table 2. Classification accuracy**

| Accuracy | Generated | Genuine |
|---|---|---|
| Mean | 18,2 % | 79,7 % |
| 95% CI | 14.5% - 21.9% | 73.7% - 85.6% |
| Highest | 27,4 % | 91,4 % |
| Lowest | 10,0 % | 58,5 % |

The difficulty of the task was brought up by some of the participants in their qualitative comments that they could write for each profile. Examples of these comments include the following: "I had to read and reread the tweet trying to understand what they were trying to say. Possibly a person, but it feels like it could be a bot." and "Once again it is impossible to tell." as well as "Too hard to tell.".

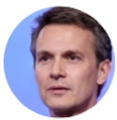

**Eugene Pohlman** @Pohlman

The Russian military has proposed a new evacuation plan for Ukrainian civilians and foreign nationals aiming to flee major cities amid Moscow's military offensive in Ukraine

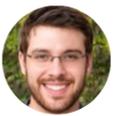

**Jeremy Ward** @JeremyWard

Zelensky says that Ukraine has gained invaluable time by Russia's obsession with Mariupol. This has allowed Ukrainian troops to retake territory north of Kyiv

**Figure 3. The least detected generated profiles**

### 4.3. Ratings of the components

As mentioned in the research design, participants were asked to rate each of the four different elements, the profile picture, post, name and handle in terms of how much it makes them suspect the account is a bot, with a slider ranging from 1 (not at all suspicious) to 10 (extremely suspicious). The results gave no real indication of any component being seen as a giveaway for either a human- or a generated account. For both classes, the mean results were ranging from 4 to 5 for all components. The highest scores were on the tweets for both genuine and generated profiles, with the mean value being 5.05 and 4.89, respectively. The lowest scoring component for both was the profile picture, with the means being 4.23 and 4.11, with the latter being the value for generated profiles. Table 3 summarizes the results and shows the 95% confidence intervals (CI) for the ratings.

We performed statistical analysis (ANOVA) to determine if there were any relationships between the ratings of the components and the classification but found no statistically significant results. Considering the complexity of the task and that the values had little variation as all four components were on typically given values between 4-5, this result is not surprising.

**Table 3. Suspiciousness of components**

| Profile | Generated | Genuine |
|---|---|---|
| Picture | 4,11 | 4,23 |
| 95% CI | 3,88 - 4,33 | 3,96 - 4,51 |
| Tweet | 4,89 | 5,05 |
| 95% CI | 4,52 - 5,25 | 4,60 - 5,49 |
| Name | 4,15 | 4,4 |
| 95% CI | 3,92 - 4,38 | 4,24 - 4,56 |
| Handle | 4,4 | 4,59 |
| 95% CI | 4,18 - 4,62 | 4,40 - 4,78 |

Rating scale: 1 (not suspicious) to 10 (extremely suspicious)

### 5. Discussion

Research question 1 asked whether humans can detect fully deep learning-generated social media profiles and posts on the feed of a social networking site. This study finds that accounts with GAN profile pictures, names drawn from a random name generator, and posts made with GPT-3 could not be distinguished from tweets and profiles created by real humans. To the best of our knowledge, this is the first time that this has been tested for a "whole" profile and not just for a generated face or text. Similar to the results of a recent experiment with GAN profile pictures alone (Nightingale & Farid, 2022), the generated profiles were viewed as more likely to be humans than the genuine human profiles. This can be explained by the fact that generated content tends to produce "average" looking data points, which in this case are the components of the profiles. At the same time, real data, i.e., the genuine profiles in this study, have more variety in components and human evaluators can make the mistake of assuming this type of noise is a sign of the generator having made an error. However, a larger sample of accounts and participants would be needed to determine if this result is generalizable or simply a result of the 18 accounts having a particular distribution of components.

The second research question, which asked whether some of the generated components can reveal that a profile is fake to a human evaluator, was left unanswered. However, the findings made in relation to RQ2 further supports the conclusion that humans are not able to distinguish real and generated profiles, as none of the generated profiles were detected by a majority of the subjects, and the ratings of the components' suspiciousness were on average very close to the central "neither nor" value.

Although the focus of this paper is not on the process of developing the fake profiles, we want to point out the accessibility and availability of the tools described in the methodology section. The ease of producing both the fake posts as well as the profile pictures was staggeringly easy. While creating and maintaining social bots would, without doubt, require intermediate to advanced programming skills, producing the components for fake profiles and posts would not, and even individuals without much training could build multiple seemingly humanlike profiles. This is primarily due to several reasons: 1) the availability of GAN-images through websites that demonstrate StyleGAN, 2) the modern tools for text generation such as GPT-3 that have no-code user interfaces, and 3) apps hosted on webpages can be used to access them even without personal access to the API. Therefore, the emergence of a growing number of realistic fake profiles is possible unless companies such as Twitter and Meta begin more actively detecting and removing profiles that, for example, have been algorithmically detected to have GAN-generated profile pictures.

These findings raise an important research impact question. What can be done to address this issue? Suppose humans cannot detect fake profiles and posts and report them manually. In that case, the role of automated detection and development of other safeguards by the companies operating the social networking sites becomes more important than before. This is due to online communities no longer being able to support moderation on their own by flagging suspicious content. Ultimately, this increases the responsibilities of social networking sites.

Moreover, one could also question whether the companies producing tools that can be used to create computational propaganda are also accountable. It should be noted that the use of GANs and text generators for malicious purposes on social media is well beyond the intentions of their respective developers and that these technologies have many potential beneficial use cases meriting their development. Moreover, companies such as OpenAI has even specifically banned using GPT-3 for the generation of offensive texts and social media content, and access to the model is terminated when infringements are caught. As evidence of this policy being enforced, during the development of this paper one of the authors had their API keys and access to the system revoked.

As a conclusion, we believe that while the availability of tools that can be used to create malicious content at scale could in theory be limited, most of the technology or alternatives to them are already published as open-source software, and thus putting the genie back into the lamp is impossible so to speak.

### 5.1. Theoretical implications

While this paper is purely an empirical study, the results can have strong implications for several theories assuming the hypothesis holds. For instance, the severity of the spiral of silence (Noelle-Neumann, 1974) could be enhanced by an influx of humanlike malicious accounts. The role of bots and fake accounts and their impact on the formation of what is the general public opinion has been studied and it has been shown by simulation that a relatively low percentage of bots can tip the discussion and trigger a spiral of silence, where a small but loud group define the perceived prevailing opinion (Ross et al., 2019). In other words, in the context of this paper, it is possible that if bad actors could create realistic-looking profiles and posts at scale, they could use them to distort the perceived public opinion.

### 5.2. Limitations

The main limitations of this study are the small number of visible components per account and the homogeneity of the profiles as all were white adult males. Moreover, the participants were mainly from a narrow demographic, as over 80% were white. These reduce the realism and generalizability of the results but were nevertheless deemed acceptable given the scope of this paper. We address the limitations with the following arguments.

First, our goal was not to determine if humans can recognize the generated profiles when given full access to the profiles and historical posts, but rather to emulate a situation where a user of a social networking site scrolls through the feed and sees multiple posts made by different users. If the profiles are not suspicious, it is unlikely that an individual would go through each profile in detail. Thus, the realistic generated profiles could pass as genuine users, and for example affect the individual's perception of what is the common opinion on a specific matter.

Second, while in a more realistic setting the experiment would have both generated and genuine profiles of various gender, ethnicity and origin as well as some with missing or less similar profile information, this would introduce too many variables that can influence the results. This could be addressed by conducting multiple experiments or introducing a significantly larger sample size. This will be addressed in our future work, which is described in the following section.

Finally, when recruiting participants, we opted not to attempt to reach a particular distribution regarding the demographics as we assumed we would in later experiments be able to make it more evenly distributed. Due to the low acceptance rate to the experiment, finding larger numbers of participants from less common demographic groups would have taken significantly more time.

### 5.3. Future work

After having conducted a pilot as well as the experiment presented in this paper, we have determined that the design of the generated profiles seems sufficient. However, the scope of the topics of the posts as well as the diversity of the profiles posting should be expanded. Previous work has suggested that GAN-generated images with non-white and female individuals are less realistic and easier to detect due to biases in the training data (Nightingale & Farid, 2022). Therefore, it would be interesting to see if this pattern remains in a richer setting where the profile pictures are accompanied by information such as a name and post.

Moreover, introducing a treatment where some participants would be given instructions on how to spot fakes could be used to determine if subjects can learn to detect fake profiles based on different components such as profile pictures or the text in the posts. This could provide valuable insights for scholars and practitioners on how to combat computational propaganda by providing users of social networking sites with appropriate instructions and training.

To reduce the possible bias of the respondents, we plan to recruit a more diverse set of participants in future experiments in both MTurk and by running the experiment using students in different regions. This will allow us to produce more robust findings as well as potentially reveal differences between groups of humans.

Lastly, to increase the realism of the simulated social media feed, the user interface will be upgraded in upcoming experiments to include more elements and the possibility to view the profile description of the accounts, as this can be done also on Twitter when hovering the cursor over a profile. This will require adding additional components for the participants to view, such as how many followers and how many accounts the profile is following, as well as the profile description, which is also known as the bio.

## 6. Conclusion

Previous experiments have shown how it is possible to create fully synthetic yet real looking pictures of faces with generative adversarial networks as well as machine generated texts, using pre-trained language models such as GPT-3 that are indistinguishable from those written by a human. In this paper, we attempted for the first time, as far as we know, to produce realistic social media profiles using these two methods to demonstrate that we have passed the point where fully generated posts and profiles can pass unnoticed by humans in a social media feed. The results of our experiment support this hypothesis as the classification accuracy was consistently low for the generated profiles. Since the generated profiles were mostly classified as genuine profiles during the experiment, we could not determine if individual components of the profiles could indicate to humans which profiles are real humans and which are generated.

However, we are careful of making strong claims or generalizing based on the results until further experiments are conducted and some of the limitations of this study are addressed. While we believe that detecting generated content and fake profiles in the feed is difficult, we hypothesize that if given access to full profiles it would be much easier for humans to spot suspicious accounts. We believe though that most humans would not go through the effort of checking each profile they come across in a feed, and thus the results of this paper can be considered concerning. Ultimately, this study suggests that making believable fake profiles with minimal human involvement is possible. Considering that fake profiles can distort online discussions and efficiently spread misinformation (Bessi & Ferrara, 2016; Shao et al., 2018), automatic detection of removal of such accounts should be the top priority of social networking sites as the end user cannot be expected to distinguish fake from real.